\documentclass[aps,twocolumn,showpacs]{revtex4-1}
\usepackage{graphicx}
\include{amssym}

\begin{document}
\title{Electromagnetic interactions of mesons induced by the axial-vector -- pseudoscalar mixings}
\author{A. A. Osipov}
\email[]{aaosipov@jinr.ru} 
\altaffiliation{}
\affiliation{Bogoliubov Laboratory of Theoretical Physics, Joint Institute for Nuclear Research, Dubna, 141980, Russia}

\author{M. M. Khalifa}
\email[]{mkhalifa1186@gmail.com}
\thanks{}
\affiliation{Moscow Institute of Physics and Technology, Russia, and Department of Physics, Al-Azhar University, Cairo, Egypt }

\begin{abstract}
It is shown that the diagonalization of the axial-vector -- pseudoscalar transitions in the effective meson Lagrangian in presence of electromagnetic interactions leads to a deviation from the vector meson dominance picture which usually arises in the Nambu - Jona-Lasinio model. The essential features of such modification of the theory are studied. Some important examples are considered in detail.
\end{abstract}

%Keywords: QCD, Chiral transformations, Spontaneous symmetry breaking, Hadronic effective field theory, Axial-vector mesons.
%\pacs{11.30.Rd, 11.30.Qc, 12.39.Fe, 14.40.-n}
\maketitle

\section{Introduction}
The electromagnetic interactions of mesons can be introduced in the corresponding chiral Lagrangian through the replacement of the usual derivatives by the gauge covariant ones \cite{Berstein68}. In particular, in the Nambu - Jona-Lasinio (NJL) model, this has been done long ago in \cite{Ebert83,Volkov83,Volkov84,Volkov86,Ebert86,Volkov86,Volkov06}. It has been shown that in the presence of vector mesons the picture appears to be identical to the vector dominance model, where the photons interact with quarks only through the exchange of $\rho^0, \,\omega$ and $\phi$ mesons.     

In this picture, however, there is a feature that is apparently, but only apparently, unrelated to the problem of electromagnetic interactions of mesons. Through the study of effective chiral Lagrangians with spin-1 mesons, it has been realized that they possess a cross term $\vec a'_\mu\partial^\mu\vec\pi$, i.e. the axial-vector $a'_\mu$ and pseudoscalar $\pi$ fields mix \cite{Gasiorovicz69,Volkov85,Osipov17,Osipov17ap,Osipov17plb}. Consequently, one should diagonalize the free part of the Lagrangian by introducing a physical axial-vector field $\vec a_\mu$. The non-diagonal term $\vec a'_\mu \partial^\mu \vec\pi$ is usually eliminated by a linearized transformation $\vec a'_\mu =\vec a_\mu +c\partial_\mu\vec\pi$ with a well-defined coupling $c$. Our observation is that in the presence of electromagnetic interactions the derivative of the pseudoscalar field, $\partial\pi$, in this conventional change of variables, should be replaced by the covariant one, ${\mathcal D}\pi$, otherwise the noncovariant diagonalization would ruin the gauge symmetry of the Lagrangian \cite{Osipov18jetp}. Somehow this obvious step is totally ignored in the literature.  

The purpose of this paper is to study the consequences of such covariant diagonalization in the photon-meson Lagrangian. Our starting point is the NJL model with $SU(2)\times SU(2)$ chiral symmetric four quark interactions (see, for instance, \cite{Osipov17ap}). Here we extend this model by including electromagnetic interactions and show that the covariant diagonalization leads to new electromagnetic vertices where a quark-antiquark pair interacts directly with the photon and the pion. As a result, the theory deviates from the vector meson dominance (VMD) scheme, but possesses the  gauge symmetry. 

To illustrate our theoretical arguments, we give several examples. The aim is to reveal the specific role of the new electromagnetic vertices induced by the $\pi a_{\mu}$-diagonalization. For instance, in the case of the $a_1\to\pi\gamma$ decay the results of the old and new approaches are shown to be identical on the mass shell. The $\gamma\pi\pi$ amplitude does not change. The anomalous $f_1(1285)\to\gamma\pi^+\pi^-$ decay amplitude is shown to be gauge invariant in both cases, but the results differ. The $a_1(1260)\to\gamma\pi^+\pi^-$ amplitude is not gauge invariant in the conventional approach, but it is invariant in the new version. The latter two processes give a very interesting and rare example for which the surface term of the triangle anomaly cannot be fixed by the Ward identities. As we will show, the amplitude contains a free parameter which should be determined from the experiment. A similar case has been studied previously in the chiral Schwinger model  [a $U(1)$ gauge field coupled to chiral fermions in two dimensions] \cite{Jackiw85,Jackiw00}.       
          
The outline of the paper is as follows. The effective quark Lagrangian with $SU(2)_L\times SU(2)_R$ chiral symmetric four-quark interactions is presented in Sec. \ref{L}. Here we introduce the auxiliary bosonic fields, add the electromagnetic interactions, and discuss shortly the evaluation of the real part of the one-loop quark determinant. We also discuss the solution of the $\pi a_\mu$ mixing problem, showing that the gauge covariant diagonalization leads to new electromagnetic vertices in the Lagrangian. This section contains the main result of our paper. In Sec. \ref{se} we give support to the correctness of the above modification of the theory by calculating different electromagnetic processes. The following examples are considered: (a) The $a_1\to\pi\gamma$ decay in Sec. \ref{seA}; (b) The $\gamma\pi\pi$ vertex in Sec. \ref{seB}; (c) In Sec. \ref{seC} the meson effective Lagrangian is used to describe the anomalous $f_1(1285)\to\gamma\pi^+\pi^-$ decay amplitude; (d) In Sec. \ref{seD} we obtain the $a_1(1260)\to\gamma\pi^+\pi^-$ decay amplitude in the one-quark-loop approximation and show how the gauge invariance is restored due to new contributions induced by the $\pi a_\mu$ covariant diagonalization. We summarize our results in Sec. \ref{concl}.  

%%%%%%%%%%%%%% SECTION 2 %%%%%  
\section{Effective Lagrangian }
\label{L}
Let us consider a system of $N_c\times N_f=3\cdot 2=6$ light Dirac quark fields $q(x)$ and an equal amount of antiquarks $\bar q(x)$ (the color and flavor indices are suppressed) with $SU(2)_V\times SU(2)_A$ chiral symmetric four-fermion interactions, and the $U(1)$ gauge invariant electromagnetic interactions. The Lagrangian density 
\begin{eqnarray}
\label{lag}
&&{\cal L}=\bar q(i\gamma^\mu{\cal D}_\mu -\hat m)q + {\cal L}_{S} + {\cal L}_{V}+{\cal L}_{em}, \\
\label{lagsp}
&&{\cal L}_{S}=(G_S/2)\left[(\bar qq)^2+(\bar qi\gamma_5\vec\tau q)^2 \right], \\
\label{lagva}
&&{\cal L}_{V}=-(G_V/2)\left[(\bar q\gamma^\mu\tau_a q)^2+(\bar q\gamma^\mu\gamma_5\tau_a q)^2 \right], \\
\label{lagem}
&&{\cal L}_{em}=-(1/4)F^{\mu\nu}F_{\mu\nu},
\end{eqnarray}
includes both spin-0 and spin-1 four-quark couplings with dimensional constants $G_S$ and $G_V$ correspondingly; $\hat m=\hat m_u=\hat m_d$ is a current quark mass; $\tau_a=(\tau_0,\vec\tau)$ for $a=0,1,2,3$, where $\tau_0$ is a $2\times 2$ unit matrix, and $\vec\tau$ are the $SU(2)$ Pauli matrices; $\gamma^\mu$ are the standard Dirac matrices in four dimensions. The covariant derivative is given by ${\cal D}_\mu =\partial_\mu -ieQA_\mu$, where the matrix $Q=1/2(\tau_3 +1/3)$ accumulates the electromagnetic charges of $u$ and $d$ quarks in relative units of the proton charge $e>0$; $A_\mu$ is a 4-potential of the electromagnetic field, and $F_{\mu\nu}=\partial_\mu A_\nu -\partial_\nu A_\mu$. 

The $U(1)$ gauge transformations 
\begin{equation}
\label{gauge}
q\to q'=e^{i\phi eQ}q, \quad A_\mu\to A'_\mu =A_\mu + \partial_\mu \phi 
\end{equation}  
are parameterized  by a local phase $\phi (x)$. The Lagrangian density ${\cal L}_{V}$ is chosen to be symmetric with respect to the $U(2)_V\times U(2)_A$ chiral transformations, because we are going to discuss the vector meson dominance mechanism which requires, in the case considered, two neutral vector meson states $\rho^0$ and $\omega$. The global transformations of the chiral group can be parameterized by eight real parameters: $\alpha_a$ and $\beta_a$. For small values of the parameters an infinitesimal change of the quark $\delta q=q'-q$ and antiquark $\delta\bar q=\bar q\,'-\bar q$ fields is given by 
\begin{equation}
\label{qt}
\delta q=i(\alpha +\gamma_5\beta )q, \quad \delta \bar q=i\bar q(-\alpha+\gamma_5\beta ),
\end{equation}
where $\alpha =\alpha_a\tau_a/2$, and $\beta =\beta_a\tau_a/2$. 

As explained by Nambu and Jona-Lasinio, the Lagrangian density ${\cal L}$ is apparently of the symmetry breaking type, in the sense that starting from some critical value of $G_S$ the minimum of the effective potential occurs for non-zero values of $\langle \bar q q\rangle\neq 0$ and the constituent quark mass $m$. This is just the chiral symmetry breaking phenomenon. In the non-symmetric vacuum, the physical spectrum contains $q\bar q$ bound states. Therefore, it is convenient to introduce the meson variables in the corresponding functional integral explicitly. This can be done by transforming the nonlinear interactions of quarks to the Yukawa type interactions of quarks with auxiliary boson fields
\begin{eqnarray}
\label{smat3}
&&S [A_\mu ]=\int [dq] [d\bar q] [ds] [d\vec p] [dv_{a\mu} ]
   [d{a}'_{a\mu} ]  \nonumber \\
&&\exp i\!\!\int\!\! d^4x\left(\bar q\, D_mq -\frac{1}{4}F_{\mu\nu}F^{\mu\nu} +\mathcal L_M\right). 
\end{eqnarray}
Here $D_m$ is the Dirac operator in the background fields
\begin{equation}
D_m = i\gamma^\mu\mathcal D_\mu -m+s+i\gamma_5 p +\gamma^\mu v_\mu
+ \gamma^\mu\gamma_5 a'_\mu .
\end{equation}
The scalar, pseudoscalar, vector and axial vector fields are $s=s\tau_0,\, p=\vec{p}\vec\tau,\, v_{\mu } =v_{a\mu }\tau_a,\, a'_\mu =a'_{a\mu }\tau_a$. $\mathcal L_M$ describes the meson mass part of the Lagrangian density
\begin{eqnarray}
\label{Mpart1}
\mathcal L_M =&-&\frac{1}{4G_S}\mbox{tr} \left[ (s -m+\hat m)^2+p^{2}\right] \nonumber \\
                        &+&\frac{1}{4G_V} \mbox{tr} \left[ v_\mu^{2}+(a'_\mu)^{2}\right].
\end{eqnarray}

The spontaneous symmetry breakdown leads to the $\vec{p} \vec{a}'_\mu$ mixing between the pseudoscalar and axial-vector fields already in the one-quark-loop approximation, i.e. in the same order at which the effective potential develops the non-symmetric ground state. To avoid the mixing one usually defines a new axial-vector field $\vec{a}_\mu$ through the replacement 
\begin{equation}    
\label{mix}
\vec{a}'_\mu=\vec{a}_\mu+\kappa m\partial_\mu \vec p, 
\end{equation}
where the constant $\kappa$ should be fixed to avoid the $\vec{p} \vec{a}_\mu$ term. This is a standard procedure which is widely used in the literature whether or not electromagnetic interactions are included. However, one can easily see that the replacement (\ref{mix}) adds to the Lagrangian density (\ref{smat3}) a Yukawa-type vertex $\kappa m\,\bar q\gamma^\mu\gamma_5 \partial_\mu p q$ which breaks gauge symmetry. Indeed, the gauge transformation of the pseudoscalar field is given by the adjoint representation of the group
\begin{equation}
\label{adj}
p\to p' =e^{i\phi eQ} p e^{-i\phi eQ}.
\end{equation}
It follows then that $\bar q\partial_\mu p q\to \bar q\partial_\mu p q+i\partial_\mu\phi e\bar q\left[Q, p \right]q$. Thus, this combination is not gauge invariant. The functional $S [A_\mu ]$ will be gauge invariant if, and only if, we use the covariant substitution
\begin{equation} 
\label{apiC}
a_\mu'=a_\mu +\kappa m \mathcal D_\mu p,  
\end{equation} 
where $\mathcal D_\mu p =\partial_\mu p -ie A_\mu  [Q, p ] $ instead of (\ref{mix}). In this case, both sides of the expression (\ref{apiC}) are transformed over the adjoint representation of the gauge group. That ensures the preservation of the gauge invariance of the functional $S [A_\mu ]$.

Our purpose now is to obtain the effective meson theory defined by the vacuum-to-vacuum transition amplitude (\ref{smat3}), where we make the replacement (\ref{apiC}). After the replacement, the differential operator $D_m$ becomes  
\begin{eqnarray}
D_m&=&i\gamma^\mu d_\mu -m+s+i\gamma_5 p, \\
d_\mu &=&\partial_\mu -i\Gamma_\mu , \\
\Gamma_\mu &=& v_\mu +eQ A_\mu +  \gamma_5 \left(a_\mu +\kappa m \mathcal D_\mu p \right).
\end{eqnarray}
This modification leads to the following consequences. Consider the replacement of variables 
\begin{equation}
\label{vdr}
v_{\mu}\to v_{\mu}-eQA_\mu
\end{equation} 
made in the functional $S [A_\mu ]$ [It is equivalent to the replacements $v_\mu^0\to v_\mu^0 - eA_\mu /6$, and $v_\mu^3\to v_\mu^3 - eA_\mu /2$]. It removes the $A_\mu$ dependence from $D_m$, except in the covariant derivative $\mathcal D_\mu p$ 
\begin{equation}
\Gamma_\mu \to \Gamma_\mu = v_\mu +  \gamma_5 \left(a_\mu +\kappa m \mathcal D_\mu p \right).
\end{equation}
In other words, when the covariant diagonalization is introduced, the direct interaction of photons with quarks does not vanish. There is still a vertex which couples the electromagnetic field with the pion and quarks. This yields a deviation from the vector meson dominance picture. The latter aspect is new [in the sense that it has never been considered before in the NJL model approach] and is the main subject of our study here.   

While there is no  direct coupling of a single photon with quarks, there may perfectly well be the couplings of the photon with the neutral vector mesons. We can see this from the mass part of the Lagrangian density, which now changes to 
\begin{eqnarray}
\label{Mpart2}
&&\mathcal L_M \to \mathcal L_M=-\frac{1}{4G_S}\mbox{tr} \left[ (s -m+\hat m)^2+p^{2}\right] \nonumber \\
&&+\frac{1}{4G_V} \mbox{tr} \left[ (v_{\mu}-eQA_\mu )^{2}+(a_\mu +\kappa m \mathcal D_\mu p)^{2}\right].
\end{eqnarray}
A typical Lagrangian of the vector meson dominance arises from the following term  
\begin{eqnarray}
\label{vmd}
&&\frac{1}{4G_V} \mbox{tr} (v_{\mu}-eQA_\mu )^{2}=\frac{m_\rho^2}{2}\left(\omega_\mu^2+(\rho^0_\mu )^2 +2\rho^+_\mu\rho^-_\mu \right) \nonumber \\
&&-\frac{e}{g_\rho}m_\rho^2 A_\mu \left(\rho^0_\mu +\frac{\omega_\mu}{3}\right)+\frac{5e^2m_\rho^2}{9g_\rho^2}A_\mu^2.
\end{eqnarray}
Here, the physical states of vector fields have been introduced [$v^0_\mu =(g_\rho /2) \omega_\mu$, $\vec v_\mu =(g_\rho /2) \vec\rho_\mu$] and the mass formula $g_\rho^2/(4G_V)=m_\rho^2$ has been used.    

Now one should integrate over the quark fields 
\begin{eqnarray}
\label{qint}
\int  [dq] [d\bar q] \exp\left( {i\!\!\int\!\! d^4x\bar qD_mq}\right)&=&\det D_m \nonumber \\
      &=& e^{\mbox{\footnotesize Tr} \ln D_m}.
\end{eqnarray}
The path integral of the Gaussian type accounts for the one-quark-loop contribution to the effective action. The result is given by the non-local functional determinant (up to an overall constant). The trace, Tr, should be calculated over color, Dirac, flavor indices and it also includes the integration over coordinates of the Minkowski space-time.

In particular, the contribution of the chiral determinant to the real part of effective action is
\begin{equation}
\label{seff}
S_{\mbox{\footnotesize eff}}= -\frac{i}{2}\,\mbox{Tr}\ln D_m^\dagger D_m =i\mathcal L_{\mbox{\footnotesize eff}}.
\end{equation}
The consistent approximation scheme to obtain from the non-local chiral determinant (\ref{qint}) the local long wavelength (low-energy) expansion for the effective action of mesons $S_{\mbox{\footnotesize eff}}$ is the Schwinger-DeWitt technique \cite{Schwinger54,DeWitt65,Ball89} (see details, for instance, in \cite{Osipov17ap}). We will restrict ourselves to the first and second-order Seeley-DeWitt coefficients. These coefficients accumulate the divergent part of the effective action, which is regularized here by the ultraviolet cutoff $\Lambda$. Let us recall that the result of such calculations is well known [in the sense that the only difference between the expression for $D_m$ obtained in \cite{Osipov17ap} and $D_m$ here is the replacement of the usual derivative $\partial_\mu p$ by the gauge covariant one $\mathcal D_\mu p$]. Thus, we can simply use that result by writing    
\begin{eqnarray}
\label{effL}
&&\mathcal L_{\mbox{\footnotesize eff}}=-\frac{\hat m}{4mG_S}\,\mbox{tr}\, (s^2+p^{2}) \nonumber \\
&&+\frac{1}{4G_V}\mbox{tr}\,\left[(v_{\mu}-eQA_\mu )^{2}+(a_\mu+\kappa m\mathcal D_\mu p\,)^2\right] 
\nonumber \\
&&+I_2\mbox{tr}\left\{(\bigtriangledown_\mu s )^2 + (\bigtriangledown_\mu p)^2 \right. \nonumber \\
&&\left. -(s^2-2ms +p^2)^2 -\frac{1}{3}(v_{\mu\nu}^2+a_{\mu\nu}^2) \right\},
\end{eqnarray}
where the factor $I_2$ is 
\begin{equation}
\label{j01}
I_2 =\frac{N_c}{(4\pi )^2}\left[\ln\left(1+\frac{\Lambda^2}{m^2}\right)-\frac{\Lambda^2}{\Lambda^2+m^2}\right], 
\end{equation}
and we adopt the following notations  
\begin{eqnarray}
\label{covder}
\bigtriangledown_\mu s &=&\partial_\mu s -\{a_\mu +\kappa m\mathcal D_\mu p, p \}, \nonumber \\
\bigtriangledown_\mu p&=&\partial_\mu p-i[v_\mu, p]+\{a_\mu +\kappa m\mathcal D_\mu p, s-m\}, \nonumber \\
v_{\mu\nu}&=&\partial_\mu v_\nu - \partial_\nu v_\mu -i[v_\mu , v_\nu ] \nonumber \\
&-&i[a_\mu +\kappa m\mathcal D_\mu p, a_\nu +\kappa m\mathcal D_\nu p ], \nonumber \\
a_{\mu\nu}&=&\partial_\mu a_\nu - \partial_\nu a_\mu -ie\kappa m F_{\mu\nu} [Q,p] \nonumber \\
&+&ie\kappa mA_\mu [Q,\partial_\nu p]-ie\kappa mA_\nu [Q,\partial_\mu p] \nonumber \\
&-&i[a_\mu +\kappa m\mathcal D_\mu p, v_\nu]-i[v_\mu,a_\nu +\kappa m\mathcal D_\nu p ]  .
\end{eqnarray}
Notice that the electromagnetic field $A_\mu $ drops out from $\bigtriangledown_\mu s$, due to the simple algebraic  properties $\{  [Q,p], p\} =[Q,p^2]=0$ [the commutator of two diagonal matrices is zero]. We have also taken into account that the antisymmetric combination $\partial_\mu a_\nu - \partial_\nu a_\mu$, after the replacement (\ref{apiC}), is changed to $\partial_\mu a_\nu - \partial_\nu a_\mu +\kappa m (\partial_\mu \mathcal D_\nu p -\partial_\nu \mathcal D_\mu p)$. In the standard case [$\mathcal D_\mu p \to \partial_\mu p$] it would not change, but the presence of the electromagnetic field leads to the three new contributions shown in our expression for $a_{\mu\nu}$.    

Some comments about formula (\ref{effL}) are still in order. To get this Lagrangian density we have used the gap equation 
\begin{equation}
\label{gap}
m-\hat m= mG_S I_1,  
\end{equation}
where 
\begin{equation}
I_1=\frac{N_c}{2\pi^2}\left[\Lambda^2 -m^2\ln \left(1+\frac{\Lambda^2}{m^2}\right)\right].
\end{equation}
It is assumed that the strength of the quark interactions is large enough, $G_S > (2\pi )^2/(N_c\Lambda^2)$, to generate a non-trivial, $m \neq 0$, solution of Eq. (\ref{gap}) [even if the current quarks would be massless].

The Lagrangian density $\mathcal L_{\mbox{\footnotesize eff}}$ does not contain $\vec p \vec a_\mu$-mixing. This is because of the cancellation which occurs between the three different contributions to the nondiagonal $\vec p\vec a_\mu$-mixing term in $\mathcal L_{\mbox{\footnotesize eff}}$. It restricts the numerical value of the parameter $\kappa$ to the following one
\begin{equation}
\label{kappa}
\frac{1}{2\kappa}=m^2+ \frac{1}{16G_V I_2}. 
\end{equation}

The free part of the Lagrangian density $\mathcal L_{\mbox{\footnotesize eff}}$ must have a canonical form. This can be done by the redefinition of the fields
\begin{eqnarray}
&& s =g_\sigma \sigma, \ \vec p=g_\pi\vec\pi, \\
&& v^0_\mu =\frac{g_\rho}{2}\omega_\mu, \ a^0_\mu =\frac{g_\rho}{2} f_{1\mu}, \\ 
&&\vec v_\mu=\frac{g_\rho}{2}\vec\rho_\mu, \ \vec a_\mu =\frac{g_\rho}{2}\vec a_{1\mu}. 
\end{eqnarray}
The renormalization constants $g_\sigma , g_\pi , g_\rho$ and masses of meson states are functions of the $I_2$ and the constant $Z^{-1}=1-2\kappa m^2$
\begin{eqnarray}
&&g_\sigma^2=\frac{1}{4I_2},\quad  g_\pi^2=Zg_\sigma^2,\quad  g_\rho^2=6g_\sigma^2, \\
&&m_\pi^2=\frac{\hat mg_\pi^2}{mG_S},\quad m_\sigma^2=4m^2+Z^{-1}m_\pi^2, \\
&&m_\rho^2=m_\omega^2=\frac{3}{8G_VI_2}, \\
&&m_{a_1}^2=m_{f_1}^2=m_\rho^2+6m^2.  
\end{eqnarray}

Apart from the language of Schwinger-DeWitt method, there is a more practical way to study the consequences of the theory (\ref{smat3}). Indeed, the vertices of the Lagrangian density $\mathcal L_{\mbox{\footnotesize eff}}$ [together with the corresponding coupling constants] can be obtained by calculating the one-quark-loop diagrams and keeping only the leading terms in the derivative expansion which dominate in the long-wavelength approximation \cite{Volkov84,Volkov86}. These direct calculations provide to be useful when one considers certain low energy processes. In the following, we will apply this method to the calculation of anomalous processes, i.e. in considering the imaginary part of the chiral determinant. 

%%%%%%%%%%%     SECTION 3      %%%%%%%%%%%%%%%%%%%%%%
\section{Some examples }
\label{se}
It should be emphasized that the transformation (\ref{apiC}) represents a change of variables in the path integral (\ref{smat3}), which does not destroy neither the chiral nor the gauge structure of the functional $S[A_\mu ]$, and, therefore, does not change the physical content of the theory. In particular, this means that the elements of the $S$-matrix on the mass surface must coincide with the results of similar approaches, where other types of $\vec p \vec a_\mu$-diagonalization are used \cite{Osipov17plb}. Such equivalence theorem is known in axiomatic field theory (Haag's theorem \cite{Haag58}), as well as in its Lagrangian version \cite{Chisholm61,Salam61}. Unfortunately, there is no basis for arguing that the replacement (\ref{mix}) is reliable for the theory described by the functional $S[A_\mu ]$. Indeed, it breaks the local gauge symmetry of $S[A_\mu ]$. Thus, the theories obtained as the result of replacements (\ref{mix}) and (\ref{apiC}) belong to the two distinct equivalence classes. The goal of this section is to make clear that these replacements lead to different physical results. 

%%%%     SUBSECTION  A
\subsection{$\mathbf a_1 \pi \gamma $-vertex}
\label{seA}
\begin{figure}
\resizebox{0.40\textwidth}{!}{%
 \includegraphics{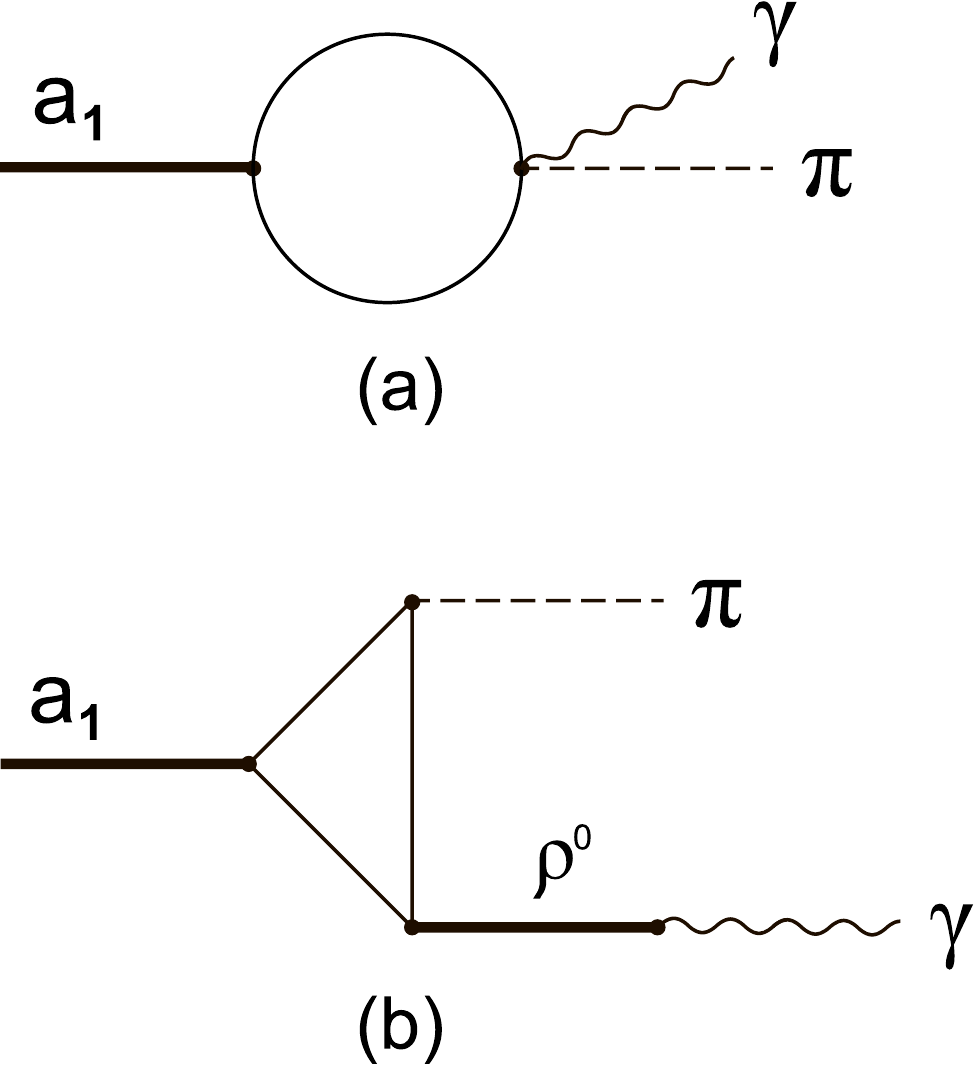}}
\caption{The typical Feynman diagrams describing the $a_1\pi\gamma$ vertex: (a) The non-VMD contribution; (b) The VMD contribution. The sum of these contributions vanishes in the leading order of the derivative expansion.}
\label{fig1}      
\end{figure}

To write an expression for $a_1\pi\gamma$-vertex, let us think of different contributions arising from the Lagrangian density (\ref{effL}).  These can be illustrated by the two Feynman diagrams, shown in Fig.\ref{fig1}. 

%% Fig.1 %%%%%%%%%%
%%%%%%%%%%%%%%%%

The diagram (a) collects the terms originated by the replacement (\ref{apiC}). Therefore, they have the non-VMD origin. These terms come out from $\mathcal L_M$, $(\bigtriangledown_\mu p)^2$, and $a_{\mu\nu}^2$ parts of $\mathcal L_{\mbox{\footnotesize eff}}$. The first two are
\begin{eqnarray}
&&\frac{1}{4G_V}\mbox{tr}\left(a_\mu +\kappa m \mathcal D_\mu p\right)^2 \to -ie\frac{\kappa m}{2G_V}\mbox{tr} \left(a_\mu [Q,p]\right) A^\mu , \nonumber \\
&&I_2\mbox{tr}(\bigtriangledown_\mu p)^2\to -8ie\kappa m^3 I_2 \mbox{tr} \left(a_\mu [Q,p]\right) A^\mu ,
\end{eqnarray}
or, after summing them, one finds
\begin{eqnarray}
\label{1p2}
&-&2ie(\kappa m)\left(\frac{1}{4G_V}+4m^2 I_2 \right) \mbox{tr} \left(a_\mu [Q,p]\right) A^\mu \nonumber \\
&=&-2ie(\kappa m)\frac{m_{a_1}^2}{g_\rho^2}\mbox{tr} \left(a_\mu [Q,p]\right) A^\mu.
\end{eqnarray}
The third term, from $a^2_{\mu\nu}$, gives  
\begin{eqnarray}
\label{three}
-\frac{I_2}{3}\mbox{tr}\, a^2_{\mu\nu}& \to& 2ie (\kappa m) \frac{1}{3}I_2\, \mbox{tr}\left\{\bar a_{\mu\nu}\left(F^{\mu\nu}[Q,p]\right.\right.\nonumber \\
&+&\left.\left. A^\nu [Q, \partial^\mu p ]-A^\mu [Q, \partial^\nu p]\right)\right\},
\end{eqnarray}
where $\bar a_{\mu\nu}=\partial_\mu a_\nu -\partial_\nu a_\mu$. 

 Combining (\ref{1p2}) and (\ref{three}), we obtain [after some redefinitions of fields] a Lagrangian density which is associated with the diagram (a) in Fig.\ref{fig1}. 
\begin{eqnarray} 
&&\mathcal L_{a_1\pi\gamma}^{(a)}= -\frac{i}{2}eg_\rho f_\pi Z\, \mbox{tr}\, \left\{ a_1^{\mu} [\mathcal A_\mu, \pi ]   \right.\nonumber \\
&& \left. - \frac{\bar a_1^{\mu\nu}}{2m_{a_1}^2}\left( [\mathcal F_{\mu\nu}, \pi ]+ [\mathcal A_\nu, \partial_\mu \pi]-[\mathcal A_\mu, \partial_\nu\pi]\right)\right\}
\end{eqnarray} 
Here, $\mathcal A_\mu =A_\mu Q$, $\mathcal F_{\mu\nu}=\partial_\mu\mathcal A_\nu -\partial_\nu\mathcal A_\mu$., and $f_\pi =m/g_\pi$ is the weak pion decay constant.  

Noting that 
\begin{eqnarray}
\bar a_1^{\mu\nu}  [\mathcal F_{\mu\nu}, \pi ]& =& -2\partial^\nu a_1^{\mu}  [\mathcal F_{\mu\nu}, \pi ] = -2\partial^\nu \left( a_1^\mu [\mathcal F_{\mu\nu}, \pi] \right) \nonumber \\
&+&2a_1^\mu \left( [\partial^\nu  \mathcal F_{\mu\nu},\pi ]   +  \mathcal F_{\mu\nu},\partial^\nu\pi ]  \right),
\end{eqnarray}
$\mathcal L_{a_1\pi\gamma}^{(a)}$ can be finally rewritten [after rearrangement of derivatives and omitting a total divergence] as follows
\begin{eqnarray} 
&&\mathcal L_{a_1\pi\gamma}^{(a)}= -\frac{i}{2}eg_\rho f_\pi Z\, \mbox{tr}\, \left\{ a_1^{\mu} [\mathcal A_\mu, \pi ]   \right.\nonumber \\
&& \left. + \frac{1}{m_{a_1}^2}\left( \mathcal F_{\mu\nu} [a_{1}^\mu , \partial^\nu \pi ]+\bar a_1^{\mu\nu} [\mathcal A_\mu, \partial_\nu\pi]\right)\right\}.
\end{eqnarray} 

The diagram (b) describes the standard VMD contribution. It is easy to see, by combining (\ref{vmd}) and (\ref{effL}), that $\mathcal L_{a_1\pi\gamma}^{(b)}= -\mathcal L_{a_1\pi\gamma}^{(a)}$. Therefore, the sum of these two diagrams vanishes. It means that there is no $a_1 \pi \gamma $-vertex in the theory described by $S[A_\mu]$ in leading order of the derivative expansion. 

While there is no $a_1 \pi \gamma $-vertex when one restricts to the first two Seeley-DeWitt coefficients in the asymptotic expansion of the theory, there may perfectly well be this vertex in the next stage of such expansion. Notice, that the approach based on the replacement (\ref{mix}) leads to the Lagrangian density $\mathcal L_{a_1\pi\gamma}^{(b)}$, which is not zero. However, it is not difficult to see that $\mathcal L_{a_1\pi\gamma}^{(b)}$ vanishes on the mass shell of the $a_1$ meson. 

%%%%     SUBSECTION  B
\subsection{$\gamma\pi\pi$-vertex}
\label{seB}
\begin{figure}
\resizebox{0.40\textwidth}{!}{%
 \includegraphics{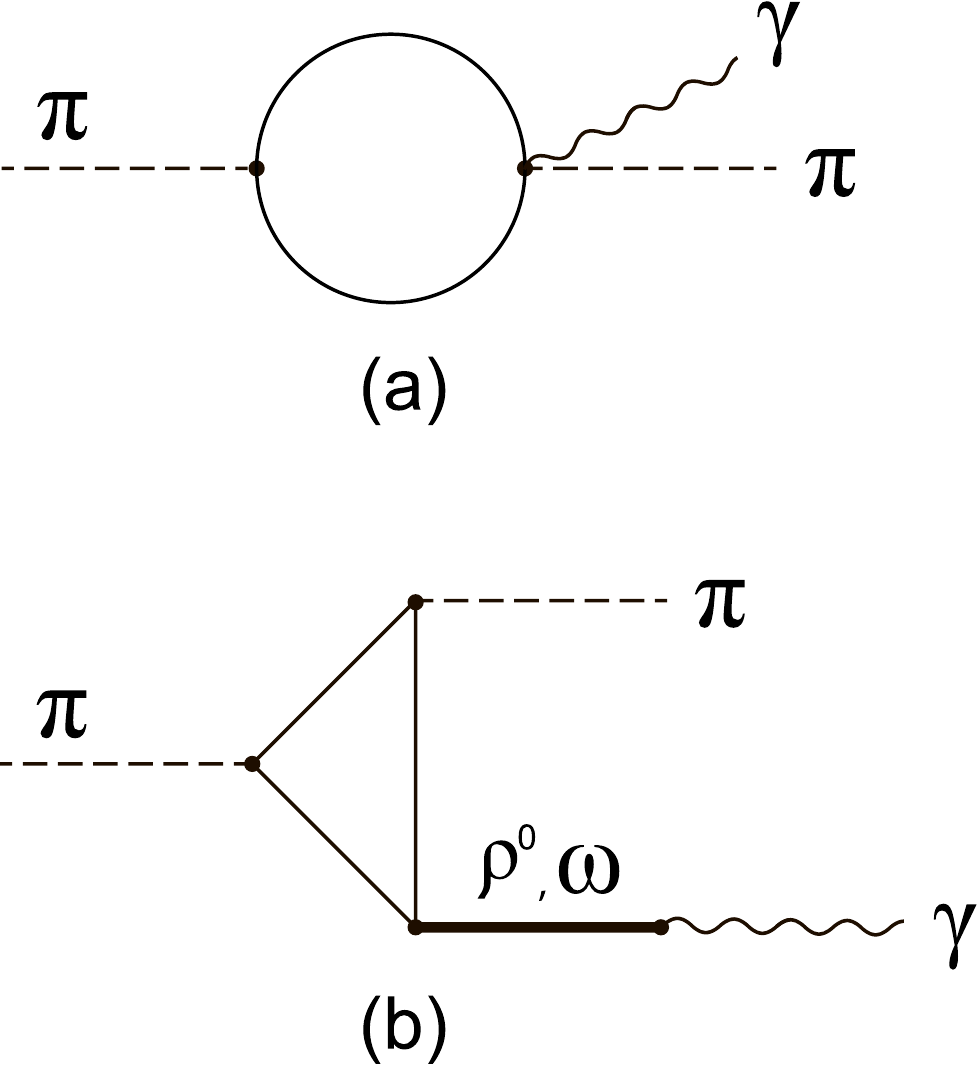}}
\caption{Two types of contributions to the $\gamma\pi\pi$ vertex: (a) The diagrams of the non-VMD origin; (b) The VMD contribution. The calculations show that the entire contribution of diagrams (a) vanishes.}
\label{fig2}      
\end{figure}

It should be appreciated that, in such a scenario, the vector meson dominance picture remains unchanged when one uses the covariant replacement (\ref{apiC}) in describing the electromagnetic form factor of the pion. To show this we calculate the contribution of diagram presented in Fig.\ref{fig2}. There are only two terms which are responsible for the non-VMD part here. These are the mass part $\mathcal L_M$, and the $(\bigtriangledown_\mu p)^2$ term in (\ref{effL}). They give
\begin{eqnarray}
&&\frac{1}{4G_V}\mbox{tr}\left(a_\mu +\kappa m \mathcal D_\mu p\right)^2 \to -ie\frac{\kappa^2 m^2}{2G_V}\mbox{tr} \left(\partial_\mu p [Q,p]\right) A^\mu , \nonumber \\
&&I_2\mbox{tr}(\bigtriangledown_\mu p)^2\to 4ie\kappa m^2 Z I_2 \mbox{tr} \left(\partial_\mu p[Q,p]\right) A^\mu .
\end{eqnarray}
The sum of these two contributions vanishes. Indeed, we have 
\begin{equation}
4ie\kappa m^2 A^\mu\mbox{tr} \left(\partial_\mu p [Q,p] \right) \left(ZI_2-\frac{\kappa}{8G_V} \right) =0,
\end{equation}
where the last step is a consequence of Eq.(\ref{kappa}).   

%% Fig.2 %%%%%%%%%%
%%%%%%%%%%%%%%%%

%%%%     SUBSECTION  C
\subsection{The anomalous $f_1\to\gamma\pi\pi$ decay}
\label{seC}
The process $f_{1}(l)\to\pi^+(p_+)+\pi^-(p_-)+\gamma (p)$ has been studied recently in \cite{Osipov18prd}. The presumed underlying theory was described by the path integral $S[A_\mu ]$, where the replacement (\ref{mix}) has been done. The amplitude got three types of contributions consistent with the vector meson dominance picture: the $\rho^0$, and $a_1$-exchanges and the direct contribution. The latter two are of our special interest here. Let us recall that the $a_1$-exchange, in the model considered, contributes as a contact interaction       
\begin{eqnarray}
\label{Ta1}
T^{(a_1)}&=&-i\frac{eg_\rho}{8\pi^2f_\pi^2}e^{\mu\nu\alpha\beta}\epsilon_\beta (l)\epsilon_\alpha^*(p) \nonumber \\ 
&\times&2\kappa m^2\left[1+(1-3a)\kappa m^2\right]l_\mu q_\nu ,
\end{eqnarray} 
where $\epsilon_\beta (l)$ and $\epsilon_\alpha^*(p)$ are the polarization vectors of the $f_1$ and the photon;  the 4-momentum $q=p_+ - p_-$. The second term in the brackets is due to the replacement (\ref{mix}). The derivative coupling $\bar q\gamma^\mu\gamma_5\partial_\mu\pi q$ makes the corresponding triangle quark diagram, $f_1 a_1\pi$, linearly divergent. This superficial linear divergence appears in the course of evaluation of the overall finite integral. Shifts in the internal momentum variable of the closed fermion loop integrals induce an arbitrary finite surface term contribution proportional to $(1 - 3a)$, where $a$ is a dimensionless constant, controlling the magnitude of an arbitrary local part \cite{Jackiw72,Jackiw00}. Observing that 
\begin{equation}
\label{algebra}
e_{\mu\nu\alpha\beta}l^\mu q^\nu =e_{\mu\nu\alpha\beta} (p^\mu q^\nu -2p_+^\mu p_-^\nu)
\end{equation} 
one sees that the term $\propto p_+^\mu p_-^\nu$ breaks gauge invariance. Thus there must be other diagrams to restore the symmetry. These are the one-quark-loop box diagrams. At leading order of the derivative expansion they give the additional contribution to the amplitude
\begin{eqnarray}
\label{f1box}
T^{(box)}&=&i\frac{eg_\rho}{8\pi^2 f_\pi^2} e^{\mu\nu\alpha\beta}\epsilon_\beta (l)\epsilon_\alpha^*(p)\left[\frac{ 
p_\mu q_\nu}{Z} \right. \nonumber \\
&-& \left. \kappa m^2 (4-\kappa m^2)p_+^\mu p_-^\nu \right].
\end{eqnarray} 

Now, one can restore the gauge symmetry of the whole amplitude by fixing the parameter $a$. The requirement is to cancel the unwanted $p_+^\mu p_-^\nu$ term of the sum $T^{(a_1)}+T^{(box)}$. It gives $a=5/12$. The rest of the sum is a gauge invariant expression
\begin{equation}
\label{T}
T^{(a_1)}+T^{(box)}=iA e^{\mu\nu\alpha\beta}\epsilon_\beta (l) \epsilon^*_\alpha (p)\, p_\mu q_\nu ,
\end{equation}
where 
\begin{equation}
A=\frac{eg_\rho}{8\pi^2 f_\pi^2}\left(\frac{2-Z}{Z}+\frac{(Z-1)^2}{8Z^2} \right).
\end{equation}
Although this is most probably the way out of the problem, the meaning of that step is not completely clear. One can argue that, in this particular case, the gauge symmetry is broken at the level of terms $\propto \kappa m^2$, but exactly at this level the gauge symmetry is explicitly broken in the Lagrangian due to the replacement (\ref{mix}).
Thus, it is not clear is it safe to use the gauge symmetry argument here. 

\begin{figure}
\resizebox{0.40\textwidth}{!}{%
 \includegraphics{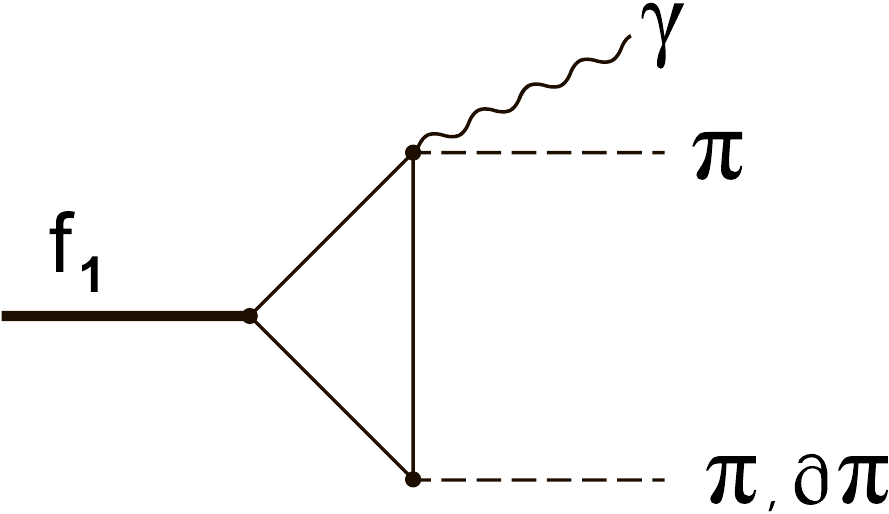}}
\caption{The Feynman diagrams describing the non-VMD contribution to the $f_{1}(1285) \to \gamma \pi^{+} \pi^{-}$ amplitude.}
\label{fig3}      
\end{figure}

What if one adds to this picture the covariant replacement (\ref{apiC})? First of all, the contribution (\ref{Ta1}) will vanish, in accord with the result of Sec. \ref{seA}. Instead, the amplitude receives new contributions from the triangle Feynman diagrams shown in Fig.\ref{fig3}, where the lower pion line ``$\pi$" represents the creation of a pion by the quark-antiquark pair due to a $\bar q\gamma_5 \pi q$ coupling and the line with ``$\partial \pi$" corresponds to a derivative coupling $\bar q\gamma^\mu\gamma_5\partial_\mu pq$. 

%% Fig.3 %%%%%%%%%%
%%%%%%%%%%%%%%%%

The first diagram [``$\pi$"] contributes to the amplitude by the following expression
\begin{eqnarray}
\label{pov}
T^{(\pi)}_\bigtriangleup&=&  \frac{ie }{f_\pi^2}N_c g_\rho (\kappa m^3) \epsilon_\beta (l)
  \epsilon^*_\alpha (p) \left[J_1^{\beta\alpha}(l,p_-) \right. \nonumber \\ 
&+&\left.  J_2^{\beta\alpha}(l,p_-) - J_1^{\beta\alpha}(l,p_+) -J_2^{\beta\alpha}(l,p_+) \right],
\end{eqnarray}
where
\begin{eqnarray}
\label{J1}
J_1^{\beta\alpha}(l,p_-)&=&\!\!\int\!\! \frac{d^4k}{(2\pi )^4}\mbox{tr}\left[ S(k,0)
      \gamma^\beta\gamma_5  S(k,l)\gamma_5 \right. \nonumber \\
&&\ \ \ \ \ \ \ \ \ \  \times\left. S(k,l-p_- )\gamma^\alpha\gamma_5 \right],      \\
J_2^{\beta\alpha}(l,p_-)&=&\!\!\int\!\! \frac{d^4k}{(2\pi )^4}\mbox{tr}\left[ S(k,-l)
      \gamma^\beta\gamma_5  S(k,0) \gamma^\alpha\gamma_5 \right.     \label{J2} \nonumber \\
&&\ \ \ \ \ \ \ \ \ \  \times\left. S(k,-l+p_- ) \gamma_5\right],      \\
S(k,l)&=& \frac{\hat k-\hat l +m}{(k-l)^2-m^2}, \quad \hat k=k_\mu\gamma^\mu.
\end{eqnarray}
In accord with the two different directions for the loop momenta we specify the loop integrals $J_1^{\beta\alpha}(l,p_-)$ and $J_2^{\beta\alpha}(l,p_-)$ by indices 1 (clockwise) and 2 (counter-clockwise). Observing that the traces are  
\begin{eqnarray}
&&\mbox{tr}[(\hat k+m)\gamma^\beta\gamma_5(\hat k-\hat l +m)\gamma_5(\hat k-\hat l +\hat p_- +m)\gamma^\alpha\gamma_5] \nonumber \\
&&=4im e_{\mu\nu\alpha\beta}(2k-l)^\mu p_-^\nu , \\
&&\mbox{tr}[(\hat k+\hat l+m)\gamma^\beta\gamma_5(\hat k +m)\gamma^\alpha\gamma_5(\hat k+\hat l -\hat p_- +m)\gamma_5] \nonumber \\
&&=-4im e_{\mu\nu\alpha\beta}(2k+l)^\mu p_-^\nu,
\end{eqnarray}
and changing in the second integral $k\to -k$, we conclude that $J_1^{\beta\alpha}(l,p_-)=J_2^{\beta\alpha}(l,p_-)$. The result then should be expanded in powers of external momenta, yielding the long wavelength approximation for the amplitude $T^{(\pi)}_\bigtriangleup \to T^{(\pi)}$
\begin{equation}
\label{Tpi}
T^{(\pi)}=-\frac{ie g_\rho}{4\pi^2 f_\pi^2}\kappa m^2 e^{\mu\nu\alpha\beta} \epsilon_\beta (l)
  \epsilon^*_\alpha (p) l_\mu (p_+ - p_-)_\nu.
\end{equation}

The contribution of the second diagram  [``$\partial\pi$"] in Fig.\ref{fig3} is
\begin{eqnarray}
\label{pov}
T^{(\partial\pi)}_\bigtriangleup&=&  \frac{ie }{f_\pi^2}N_c g_\rho (\kappa m^2)^2 \epsilon_\beta (l)
  \epsilon^*_\alpha (p) \left[I_1^{\beta\alpha}(l,p_-) \right. \nonumber \\ 
&+&\left.  I_2^{\beta\alpha}(l,p_-)- I_1^{\beta\alpha}(l,p_+) -I_2^{\beta\alpha}(l,p_+) \right],
\end{eqnarray}
where
\begin{eqnarray}
\label{I1}
I_1^{\beta\alpha}(l,p_-)&=&\!\!\int\!\! \frac{d^4k}{(2\pi
                          )^4}\mbox{tr}\left[ S(k,0)
                          \gamma^\beta\gamma_5  S(k,l)  \hat p_-
                          \gamma_5 \right. \nonumber \\
  &&\ \ \ \ \ \ \ \ \ \  \times\left. S(k,l-p_- )\gamma^\alpha\gamma_5 \right],      \\
I_2^{\beta\alpha}(l,p_-)&=&\!\!\int\!\! \frac{d^4k}{(2\pi
                          )^4}\mbox{tr}\left[ S(k,0)
                          \gamma^\beta\gamma_5  S(k,l) \gamma^\alpha\gamma_5 
                          \right.     \label{I2} \nonumber \\
  &&\ \ \ \ \ \ \ \ \ \  \times\left. S(k,p_- )\hat
     p_-\gamma_5\right].     
\end{eqnarray}
Just like in the previous case one can show [by the corresponding replacements in one of the integrals] that  $I_1^{\beta\alpha}(l,p_-)=I_2^{\beta\alpha}(l,p_-)$. However, unlike the previous case, these integrals are superficially linearly divergent even though an eventual evaluation yields a finite answer. Owing to the linear divergence, shifting the integration momentum in the closed loop changes the value of the integral, so that there is an essential ambiguity in (\ref{I1}) and (\ref{I2}). As a result, at low momenta we find $T^{(\partial\pi)}_\bigtriangleup \to T^{(\partial\pi)}$
\begin{eqnarray}
\label{Tdpi}
T^{(\partial\pi)}=&-&\frac{ie g_\rho}{4\pi^2 f_\pi^2}(\kappa m^2)^2 e^{\mu\nu\alpha\beta} \epsilon_\beta (l)
  \epsilon^*_\alpha (p) \nonumber \\ 
  &\times& (3c-2l)_\mu (p_+ - p_-)_\nu,
\end{eqnarray}
where $c_\mu$ is a free 4-vector, which, in general, can be written as a linear combination of three independent 4-vectors, entering the triangle diagram, i.e.
\begin{equation}  
c_\mu = a p_\mu +b(p_+ +p_-)_\mu  +c(p_+ -p_-)_\mu 
\end{equation}
Inserting $c_\mu$ into Eq.(\ref{Tdpi}) and taking into account (\ref{f1box}), and (\ref{Tpi}) we find the sum $T=T^{(\pi)}+T^{(\partial\pi)}+T^{(box)}$ 
\begin{eqnarray}
\label{sumf1}
&&T=\frac{ie g_\rho}{8\pi^2 f_\pi^2} e^{\mu\nu\alpha\beta} \epsilon_\beta (l)\epsilon^*_\alpha (p)  
         \left\{ \left[ (1-2\kappa m^2)^2   \right. \right.     \nonumber \\ 
&& \left.\left. - 6a\kappa^2 m^4 \right] p_\mu q_\nu  + \kappa^2 m^4 (12b-7)p_{+\mu} p_{-\nu} \right\}.  
\end{eqnarray}
When gauge invariance is enforced $(b=7/12)$, this amplitude still contains an ambiguity in the form of the undetermined constant $a$. 

To conclude this section, we will compare our result with the one obtained on the basis of the standard replacement, i.e. with the formula (\ref{T}).  One can see that the gauge invariant approach changes the result essentially. The requirement of gauge invariance, which fixes the ambiguity in (\ref{T}) by insisting that this symmetry is preserved, leads to a definite value for the constant $A$ [see Eq.(\ref{T})]. This is not the case in the consistent approach to the gauge symmetry. The formal gauge invariance of the model does not fix $A$. It means that the constant $a$ should be fixed from the experiment. It is interesting to note that the result (\ref{T}) arises from the formula (\ref{sumf1}) at $a=b=7/12$. 

%%%%     SUBSECTION  D
\subsection{The anomalous $a_1\to\gamma\pi\pi$ decay} 
\label{seD}
The calculation of the decay amplitude $a_{1}(l) \to \gamma (p) +\pi^{+} (p_+) + \pi^{-}(p_-)$, where $l, p, p_+, p_-$ are the 4-momenta of the corresponding particles, can be carried out, in the standard approach, in a similar way as was being done for the $a_{1}(1260) \to \omega \pi^{+} \pi^{-}$ decay in \cite{Osipov18hep}. One should only turn on the vector meson dominance conversion $\omega\to\gamma$, described by the Lagrangian density (\ref{vmd}). The amplitude will accumulate contributions from three different processes: (a) the $\rho^0$ exchange channel $a_1\to\gamma\rho^0\to\gamma\pi^+\pi^-$; (b) the $\rho^\pm$ exchange $a_1\to\pi^{\pm} \rho^{\mp}\to\pi^+\pi^-\gamma$; and (c) the direct decay mode $a_1\to\gamma\pi^+\pi^-$ described by the quark box diagram. 

The creation of the photon in the exchange channels is a result of the anomalous processes $a_1\to\gamma\rho^0$ and $\rho^{\pm}\to\pi^\pm\gamma$. There is no problem in evaluating these amplitudes, which are known to be gauge invariant. In view of this it seems worthwhile to concentrate on the study of the direct channels amplitudes. The calculation of the Feynman box diagrams and the separation of leading terms in the expansion in external momenta (the long-wave expansion of the fermion determinant) leads to an amplitude \cite{Osipov18hep}.
\begin{eqnarray}
\label{boxa1}
{\mathcal T}_{box}&=&i\frac{g_\rho eN_c}{8\pi^2 f_\pi^2} e^{\mu\nu\alpha\beta}\epsilon_\beta (l)\epsilon_\alpha^*(p)\left[(1-2\kappa m^2)p_\mu q_\nu  \right. \nonumber \\
&+& \left. (\kappa m^2)^2 p_+^\mu p_-^\nu \right],
\end{eqnarray} 

%% Fig.4 %%%%%%%%%%
\begin{figure}
\resizebox{0.40\textwidth}{!}{%
 \includegraphics{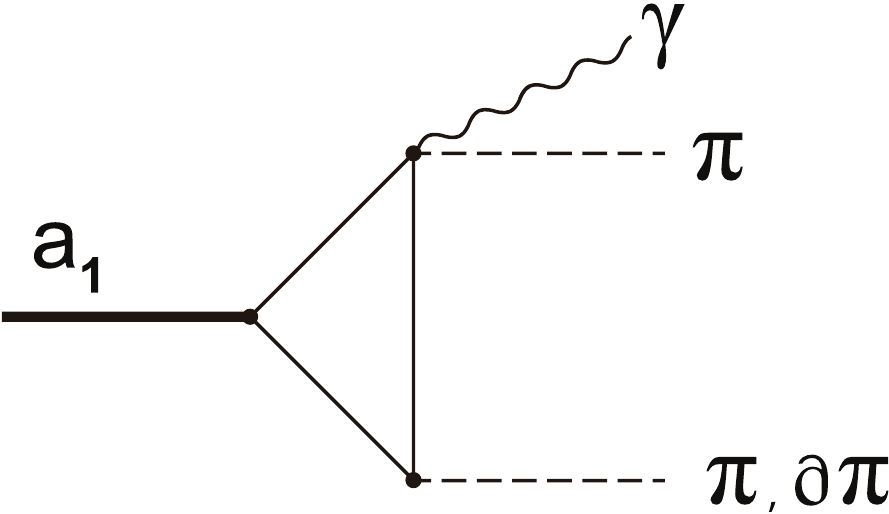}}
\caption{The typical Feynman diagrams describing the non-VMD contribution to the $a_{1}(1260) \to \gamma \pi^{+} \pi^{-}$ amplitude.}
\label{fig4}      
\end{figure}
%%%%%%%%%%%%%%%%

There is an obvious, troublesome question. If all other contributions to the amplitude are gauge invariant, how does one deal with the last term of (\ref{boxa1}) which breaks the gauge symmetry? The answer to this question cannot be found in the conventional approach. However, the consideration based on a covariant derivative (\ref{apiC}) solves the problem. Indeed, it leads to additional contributions shown in Fig. \ref{fig4}. Both anomalous triangle diagrams are finite. A single pion vertex of the first diagram is described by the Lagrangian density ${\mathcal L}_{\pi}=ig_\pi\bar q\gamma_5\pi q$. In the second diagram, this vertex is replaced by the axial-vector coupling ${\mathcal L}_{\partial\pi}=\kappa m g_\pi\bar q\gamma^\mu\gamma_5\partial_\mu\pi q$. 

Let us first write a formal expression for the amplitude with the vertex ${\mathcal L}_{\pi}$. 
\begin{eqnarray}
\label{pova1}
{\mathcal T}^{(\pi)}_\bigtriangleup&=&  \frac{ie }{f_\pi^2}N_c g_\rho (\kappa m^3) \epsilon_\beta (l)
  \epsilon^*_\alpha (p) \left[J_1^{\beta\alpha}(l,p_-) \right. \nonumber \\ 
&-&\left.  J_2^{\beta\alpha}(l,p_-) + J_1^{\beta\alpha}(l,p_+) -J_2^{\beta\alpha}(l,p_+) \right],
\end{eqnarray}
where $J_1^{\beta\alpha}(l,p_-)$ and $J_2^{\beta\alpha}(l,p_-)$ are given by Eqs. (\ref{J1}) and (\ref{J2}). It is clear that this amplitude vanishes. Indeed, due to the property $J_1^{\beta\alpha} (l,p_-) = J_2^{\beta\alpha}(l,p_-)$, the second term cancels the first one, and the fourth term cancels the third one, giving ${\mathcal T}^{(\pi)}_\bigtriangleup =0$.    

Thus, it remains to consider the contribution of the second diagram which can be written as 
\begin{eqnarray}
\label{pova2}
{\mathcal T}^{(\partial\pi)}_\bigtriangleup&=&  \frac{ie }{f_\pi^2}N_c g_\rho (\kappa m^2)^2 \epsilon_\beta (l)
  \epsilon^*_\alpha (p) \left[I_1^{\beta\alpha}(l,p_-) \right. \nonumber \\ 
&-&\left.  I_2^{\beta\alpha}(l,p_-)+ I_1^{\beta\alpha}(l,p_+) -I_2^{\beta\alpha}(l,p_+) \right],
\end{eqnarray}
where $I_1^{\beta\alpha}(l,p_-)$ and $I_2^{\beta\alpha}(l,p_-)$ are given by the formulas (\ref{I1}) and (\ref{I2}). If it were possible to shift the integration variable in these expressions, the first term would cancel the second one, and the third term would cancel the fourth one in the square brackets (\ref{pova2}) and we would obtain that ${\mathcal T}^{(\partial\pi)}_\bigtriangleup =0$. However, due to the formal linear divergence of these integrals, which is present in (\ref{I1}) and (\ref{I2}) even after traces are calculated, the surface terms arise \cite{Jackiw72}. The latter renders the result to be different from zero
\begin{equation}
 I_1^{\beta\alpha}(l,p_-)-I_2^{\beta\alpha}(l,p_-)=\frac{1}{8\pi^2}
e^{\mu\nu\alpha\beta} c_\mu  (p_-)_\nu ,
\end{equation} 
where $c_\mu$ is an arbitrary 4-momentum.

As a result, the amplitude receives a finite contribution    
\begin{equation}
\label{povterm}
{\mathcal T}^{(\partial\pi)}_\bigtriangleup =i\frac{e g_\rho N_c }{8\pi^2f_\pi^2}  (\kappa m^2)^2 \epsilon_\beta (l)
  \epsilon^*_\alpha (p) e^{\mu\nu\alpha\beta} c_\mu (p_+ + p_-)_\nu .
\end{equation}
Notice, that this is the complete result for this triangle digram. We got it without using the derivative expansion.     

The 4-vector $c_\mu$ can be represented as a linear combination of three independent momenta that are directly related to the process under consideration
$$ 
c^\mu = a p^\mu + b(p_+ - p_-)^\mu + c (p_+ + p_-)^\mu. 
$$
In fact, only two of them survive after substituting this expression in (\ref{povterm}). Consequently, the contribution of the second diagram shown in Fig. \ref{fig4} takes the form
\begin{eqnarray}
\label{pt}
{\mathcal T}^{(\partial\pi)}_\bigtriangleup&=&i\frac{e g_\rho N_c }{8\pi^2f_\pi^2}  (\kappa m^2)^2 \epsilon_\beta (l)
  \epsilon^*_\alpha (p) e^{\mu\nu\alpha\beta} \nonumber \\
&& \left[a p_\mu (p_+ + p_-)_\nu +2b(p_+)_\mu (p_-)_\nu \right].
\end{eqnarray}
One immediately sees now that choosing $b = -1/2$ one vanishes the terms that violate the gauge invariance in the sum of (\ref{boxa1}) and (\ref{pt}).
\begin{eqnarray}
\label{sum}
&&{\mathcal T}_{box}+{\mathcal T}^{(\partial\pi)}_\bigtriangleup=i\frac{g_\rho eN_c}{8\pi^2 f_\pi^2}
             e^{\mu\nu\alpha\beta}\epsilon_\beta
             (l)\epsilon_\alpha^*(p) p_\mu  \nonumber \\
&&\left[(1-2\kappa m^2) q_\nu  +  a  (\kappa m^2)^2  (p_+ + p_-)_\nu \right].
\end{eqnarray} 
The only uncertain quantity in an expression (\ref{sum}) is the constant $a$, which cannot be fixed by the vector Ward identities. 

Thus, we obtain a finite gauge invariant result, but the theory does not allow us to calculate the constant $a$. It must be fixed from the experiment. Let us recall that a similar situation occurs in a soluble two-dimensional chiral model of Schwinger. That case was analyzed in detail in \cite{Jackiw85,Jackiw00}. Here, we have discussed the four-dimensional example, which is interesting not only from the pure theoretical point of view, but also because the issue can be studied experimentally. \\

%%%%%%%%       CONCLUSIONS       %%%%%%%%%%%%
\section{Conclusions}
\label{concl}
The purpose of this paper has been to check the consistency of assuming that in the NJL model with vector mesons the procedure of $\pi a_1$-diagonalization, in the presence of electromagnetic interactions, should be carried out in a gauge covariant way. This fact is unreasonably ignored in the literature. Since the covariant derivative contains an electromagnetic field, direct interactions of a photon with a pseudoscalar meson and a quark-antiquark pair appear in the theory. This brings the theory beyond the generally accepted picture of vector meson dominance.  We have explicitly demonstrated that there are physical consequences of such a step.  

To show this, we have obtained the effective meson Lagrangian with an approximate $SU(2)\times SU(2)$ chiral symmetry, and have studied some electromagnetic processes where novel vertices are involved. The aim of providing these examples is not to offer an exhaustive overview of the possible physical consequences, but rather some examples to convince the reader that such consequences really take place. 

Note that the changes are mainly related with a modification of a local replacement of variables in the theory [instead of (\ref{mix}) we use (\ref{apiC})] and, in accord with the Chisholm's theorem \cite{Chisholm61,Salam61}, should not alter the S-matrix. It is easy to understand why, in spite of this expectation, the results differ. The reason for this is contained in the gauge symmetry requirement. Violating the gauge symmetry, the change (\ref{mix}) leads to the contradiction with the Ward identities and because of that cannot be considered as an equivalent transformation of the theory. Nonetheless, in some cases, the replacement (\ref{apiC}) leads to the same result as the replacement (\ref{mix}) [the $\gamma\pi\pi$ vertex], or the results differ by their off shell behaviour [the $a_1\to\pi\gamma$ decay]. 

The real physical consequences we have found are related with the anomalous $f_{1}(1285) \to \gamma \pi^{+} \pi^{-}$, and $a_{1}(1260) \to \gamma \pi^{+} \pi^{-}$ decays. In both cases, the new coupling $\bar qq \gamma\pi$ not only restores the local gauge symmetry, but also generates a surface contribution to the amplitude. It gives us one of the rare nontrivial field-theoretical examples of how, when calculating the final contributions from single-loop quark diagrams, there arises a surface term whose dimensionless constant can not be fixed by the theory.

%%%%%%%    ACKNOWLEDGMENTS
\section*{Acknowledgments}
A. A. O. wishes to thank M. K. Volkov for useful discussions and his interest to the work, and B. Hiller for a careful reading of the manuscript and useful correspondence.   

%\bibliography{mybibfile}
{}
\end{document}